\UseRawInputEncoding
\documentclass[prl,twocolumn,floatfix,superscriptaddress]{revtex4}

\usepackage{graphicx}
\usepackage{amsmath}
\usepackage{xcolor}
\usepackage[normalem]{ulem}
\usepackage[T1]{fontenc}
\usepackage{float}
\usepackage{comment}
\usepackage{physics}
\usepackage{natbib}

\newcommand{\ybion}{$^{171}\textrm{Yb}^+$}
\newcommand{\sstate}{\ket{2\textrm{S}_{1/2},F=0}}
\newcommand{\dstate}{\ket{2\textrm{D}_{3/2},F=2}}
\newcommand{\sz}{\ket{2\textrm{S}_{1/2},F=0,m_{\textrm{F}}=0}}
\newcommand{\dz}{\ket{2\textrm{D}_{3/2},F=2,m_{\textrm{F}}=0}}
\newcommand{\dm}{\ket{2\textrm{D}_{3/2},F=2,m_{\textrm{F}}=-1}}
\newcommand{\mF}{m_{\textrm{F}}}
\newcommand{\mic}{\,$\mu$m\:}
\newcommand{\mics}{\,$\mu$s\:}

\begin{document}

\preprint{APS/123-QED}

\title{Multi-site Integrated Optical Addressing of Trapped Ions}

\author{Joonhyuk Kwon}
\email{jookwon@sandia.gov}
\affiliation{Sandia National Laboratories, Albuquerque, New Mexico 87185, USA}
\author{William J. Setzer}
\affiliation{Sandia National Laboratories, Albuquerque, New Mexico 87185, USA}
\author{Michael Gehl}
\affiliation{Sandia National Laboratories, Albuquerque, New Mexico 87185, USA}
\author{Nicholas Karl}
\affiliation{Sandia National Laboratories, Albuquerque, New Mexico 87185, USA}
\author{Jay Van Der Wall}
\affiliation{Sandia National Laboratories, Albuquerque, New Mexico 87185, USA}
\author{Ryan Law}
\affiliation{Sandia National Laboratories, Albuquerque, New Mexico 87185, USA}
\author{Matthew G. Blain}
\affiliation{Sandia National Laboratories, Albuquerque, New Mexico 87185, USA}
\affiliation{\textrm{Present address}: Quantinuum LLC, 303 S Technology Ct, Broomfield, CO 80021, USA}
\author{Daniel Stick}
\affiliation{Sandia National Laboratories, Albuquerque, New Mexico 87185, USA}
\author{Hayden J. McGuinness}
\email{hmcgui@sandia.gov}
\affiliation{Sandia National Laboratories, Albuquerque, New Mexico 87185, USA}

\date{\today}

\begin{abstract}
 
One of the most effective ways to advance the performance of quantum computers and quantum sensors is to increase the number of qubits or quantum resources in the system. A major technical challenge that must be solved to realize this goal for trapped-ion systems is scaling the delivery of optical signals to many individual ions. In this paper we demonstrate an approach employing waveguides and multi-mode interferometer splitters to optically address multiple \!\ybion\medspace  ions in a surface trap by delivering all wavelengths required for full qubit control. Measurements of hyperfine spectra and Rabi flopping were performed on the E2 clock transition, using integrated waveguides for delivering the light needed for Doppler cooling, state preparation, coherent operations, and detection. 
We describe the use of splitters to address multiple ions using a single optical input per wavelength and use them to demonstrate simultaneous Rabi flopping on two different transitions occurring at distinct trap sites. This work represents an important step towards the realization of scalable integrated photonics for atomic clocks and trapped-ion quantum information systems.

\end{abstract}
\maketitle

Ions stored in RF Paul traps \cite{Paul:1953} have provided an effective platform for quantum information due to their stability \cite{wu:2021}, long coherence times \cite{wang:2021}, and high fidelities \cite{clark:2021}.
In addition to computing \cite{srinivas:2021,leung:2018}, ion traps have also been used for atomic clocks \cite{brewer:2019a, Burt:2021Nat}, quantum networks \cite{moehring:2007, Nichol:2022Nat}, quantum sensors \cite{marciniak:2022}, and fundamental science \cite{Zhang:2018NC, Pinkas:2023NP}. 
The development of surface traps \cite{chiaverini:2005,blain:2021} furthered the potential for advancing these applications by enabling scaling to more ions.
Using microfabrication capabilities originally developed by the semiconductor industry, surface ion traps were fabricated that could support many-ion quantum computers \cite{pino:2021, noel:2022}. 
These traps also provided a convenient platform for integrating other electronic and optical technologies \cite{moody:2022}, which are necessary for the individual addressing and readout of growing numbers of ions. 
These technologies include photonic waveguides, waveguide splitters \cite{Vasques:2022PRL}, grating couplers \cite{mehta:2016,niffenegger:2020,mehta:2020,ivory:2021}, detectors \cite{setzer:2021, Reens:2022PRL,todaro:2021}, and modulators \cite{hogle:2023}, all of which combined can support quantum systems with many ions as well as applications which require low size, weight, and power (SWaP). 

While many of these proof-of-principle experiments focused on the operation of a single ion site, simultaneous independent control of multiple ions at several trap sites had until now not been realized. Here, we demonstrate a room temperature ion trap and simultaneous optical addressing of three \ybion\medspace ions in independent wells, using only light delivered through waveguides. The light addressing two of the three ions comes from a single split input, forming a multi-ion ensemble.
Multi-mode interferometer (MMI) splitters are employed  to equally divide the light from the input waveguide into two separate waveguides, which are routed to output couplers individually addressing the trapped ions. Other splitting ratios are possible by varying the interaction length in the MMI splitters, as are other fanout ratios (e.g. 1$\times$N). This technique could be employed to deliver light to a much larger number of ions, ultimately limited by the power handling capacity of waveguides.
Large collections of ions are particularly interesting for atomic clocks, where it has been shown that separately interrogated ensembles of ions can achieve sensitivities that scale as $(\alpha N)^{-m/2}$, where $N$ is the number of ions in each of $m$ ensembles, and $\alpha$ is a protocol-dependent constant \cite{borregaard:2013}.

In this experiment we performed hyperfine spectroscopy and Rabi flopping on the E2 clock transition, consisting of Doppler cooling, state preparation, coherent operations, and detection. Only the photoionization laser beams were delivered via free space. This required multiple wavelengths ranging from UV (369 nm) to NIR (935 nm). We demonstrated the bluest wavelength used for integrated optical addressing to date, overcoming challenges related to the lithography of the grating couplers and propagation losses of the waveguides. Additionally, we measured simultaneous Rabi flopping for two different transitions occurring at different sites, which is a key capability for quantum systems that require magnetic field calibration. 
We also investigated optical cross-talk between sites. 

\section{Results}
\subsection{Experimental setup}

\begin{figure*}[t!]
\centering
    \includegraphics[width=1.8\columnwidth]{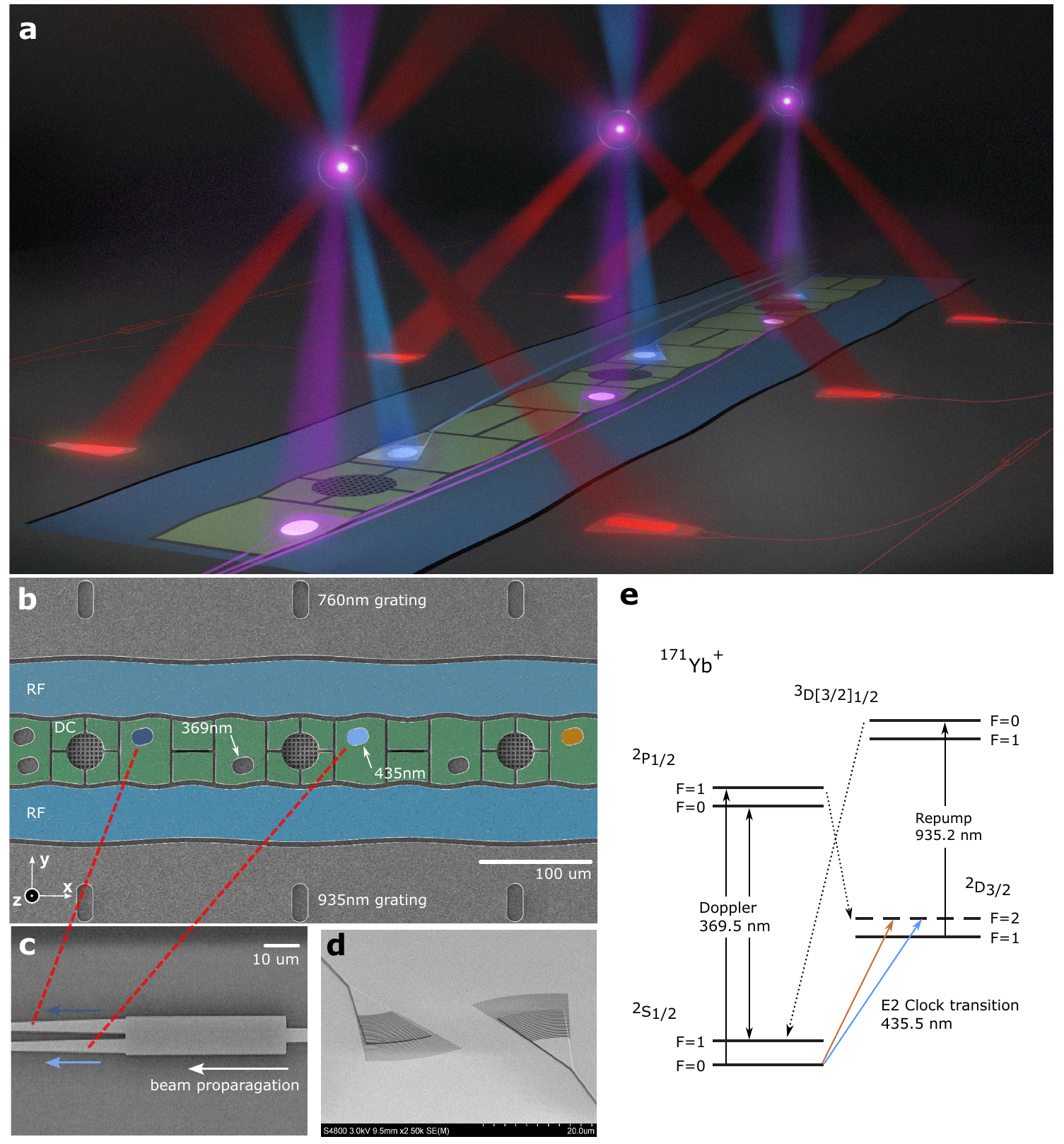}
    \caption{\textbf{Experimental setup}. \textbf{a,}  Conceptual illustration of a surface trap with ions and integrated beams. \textbf{b,} SEM image of the trapping region of the micro-fabricated surface trap, consisting of multi-color waveguide output couplers (oval structures within RF rails), electrodes (green) and RF rails (blue). 
    Single photon avalanche diodes (SPADs) are located at the center of trapping sites and are shielded by a grounded metal mesh to protect the ion from electric field perturbations when the SPADs are operational (see Methods). \textbf{c,} SEM image for an MMI splitter that routes light to waveguide output couplers (indicated in blue \& dark blue). Similar splitters are also applied for other colors. \textbf{d,} SEM image of grating output couplers. \textbf{e,} \ybion\medspace energy level diagram, indicating the three different integrated wavelengths used in the experiment.}
\label{FIG:SystemIntro}
\end{figure*}

Fig.\ \ref{FIG:SystemIntro}\textbf{b} shows a false-color scanning electron microscope (SEM) image of multiple sites in the microfabricated surface trap. Each site was optically addressed by light from four output couplers, allowing for the necessary state manipulation needed for a clock based on \ybion\!. An energy level diagram of the relevant levels is shown in Fig.\ \ref{FIG:SystemIntro}\textbf{e}. 

Ions were trapped 50\mic above the surface with a 200\mic spacing between trap sites. Light was coupled into the waveguides through input gratings located far from the trapping region and routed to the output gratings. Single-layer aluminium oxide waveguides were used for the UV light at 369 nm and 435 nm, while silicon nitride waveguides were used for 760 nm and 935 nm. These materials were chosen for their low-loss at these wavelengths and for ease of fabrication. Light from three output couplers converged at each of three trap sites along the trap axis, allowing for three ions to be cooled and manipulated using only light from the integrated photonics.

In the case of the 435 nm beams, 2.0 mW was applied to the input grating for ensemble 2 (single) and 4.3 mW to the input grating for ensemble 1 (split).
There were significant losses at the inputs that reduced the amount of light that was coupled and propagated through roughly 5 mm of the waveguides before being emitted through the output gratings (Fig.\ \ref{FIG:SystemIntro}\textbf{d}), ultimately resulting in $\pi$-times of $130$\mics\! or longer. Based on this $\pi$-time and the input power we estimate a 49 dB loss between the free-space beam launch and the ion, significantly higher than the 20 dB we measured outside the chamber. We believe that this discrepancy primarily results from additional input coupling loss, which can be improved through the implementation of edge-coupling to fibers. A more detailed description is provided in the Methods section.

Ions were trapped at room temperature with a vacuum pressure of $\approx$1$\times 10^{-11}$ Torr. A magnetic field of 5.5 G ($1.4\hat{\textrm{\textbf{x}}}+0.7\hat{\textrm{\textbf{y}}}+5.3\hat{\textrm{\textbf{z}}}$) was applied to optimize both ion fluorescence and coupling to the quadrupole transition. This bias magnetic field also broke the degeneracy of the Zeeman sublevels of the excited states.

A defining feature of these integrated photonics is that the waveguides for two of the three trap sites were fed via a common input. For each wavelength, an integrated splitter, as shown in Fig.\ \ref{FIG:SystemIntro}\textbf{c}, was used to equally divide the input light into two waveguide channels which were routed to the output gratings of the two sites. The MMI splitters were measured to have low optical losses of $<0.2$ dB per splitter. Integrated splitters allow for the control of numerous ions at different sites sites via a single input, providing a path for supporting many-ion ensembles with only a few optical inputs.

\begin{figure*}[t!]
\centering
    \includegraphics[width=1.95\columnwidth]{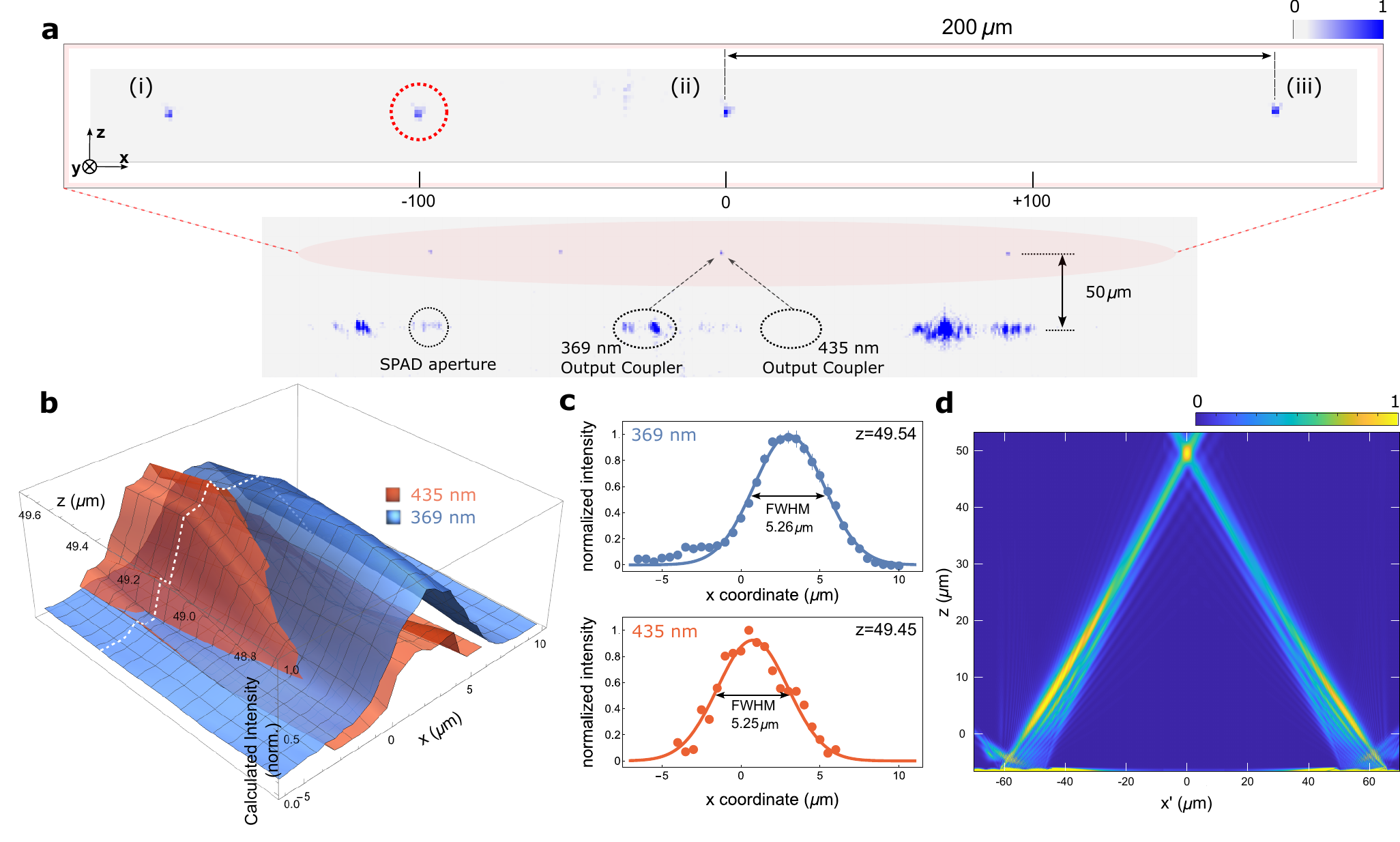}
    \caption{\textbf{Multi-site optical addressing and integrated beam profiles}.
    \textbf {a,} Image of multiple trapped ions taken with a CCD camera from the side of the trap. The top image shows three  ions illuminated by light from waveguides (i, ii, and iii), with the left two (i and ii) using light split from single sources corresponding to the different required wavelengths. The fourth ion highlighted with the dashed red circle was illuminated with free-space beams at the loading site. \textbf{b,} Beam profiles for the waveguide-delivered 369 nm and 435 nm beams. The intensities were calculated using the ion fluorescence (369 nm beam) and Rabi frequency (435 nm beam). A 3D contour plot shows normalized intensity along the x-z plane. The $x$-axis origin is at the center of the middle SPAD, consistent with the coordinates in part \textbf{a}. The beams intersected at 50\mic from the surface and near the center of the SPAD. The white-dashed line represents RF null position on z-axis.
    \textbf{c,} Measured profiles along the trap axis of the 369 nm and 435 nm at the height corresponding to the optimized interaction strength (used in Fig.\ \ref{FIG:3rabi}). Error bars show the standard error of the mean. 
    \textbf{d,} Simulated waveguide profiles for Doppler (left) and clock transition (right) beams. The $x^{\prime}$-axis in this figure is along the line between waveguide gratings. For the $z$-axis, 0 is the trap surface.
    }
\label{FIG:waveguidedetail}
\end{figure*}

\subsection{Multi-ion trapping and waveguide characterization}

State detection is particularly challenging in surface traps using integrated waveguides because light emitted from output couplers can scatter into the imaging system and overwhelm the ion fluorescence. In a standard detection scheme \cite{Streed:2011PRL,niffenegger:2020}, ions fluoresce at the same wavelength as the detection laser, rendering wavelength filters too broad to distinguish between the two. The light output from a grating is highly polarized and therefore polarization filtering is feasible, but provided insufficient filtering of scattered light in this experiment.

In addition to a conventional imaging system with both a CCD camera (Andor Zyla) and photo-multiplier tube (PMT, Hamamatsu H10682-210) for collecting fluorescence normal to the trap surface, a side-imaging lens was added to collect light along the $y$-axis. This geometry avoids collecting the off-axis scattered photons from the cooling/detection beams that emerge at $45^{\circ}$ from the waveguide output couplers, which otherwise overwhelms the ion fluorescence signal when imaged from above.
As a standard 6-inch octagonal vacuum chamber was used, the distance from the ion to the objective placed at a side viewport was larger by a factor of 5 compared to conventional overhead imaging, and therefore was much less efficient at collecting photons due the reduced numerical aperture (see Methods).
With the conventional overhead imaging system and free-space light delivery, the total count rate on the PMT was 30 kcps (kilocounts per second), with negligible background noise ($<$0.1 kcps). For waveguide-delivered light and overhead imaging, the signal-to-noise ratio was $\ll$1. Using side-imaging and waveguide-delivered light, the total count rate was 5-8 kcps with high but manageable background noise ($\sim$2 kcps) from scattered photons.
 
\begin{figure*}[t!]
\centering
    \includegraphics[width=1.95\columnwidth]{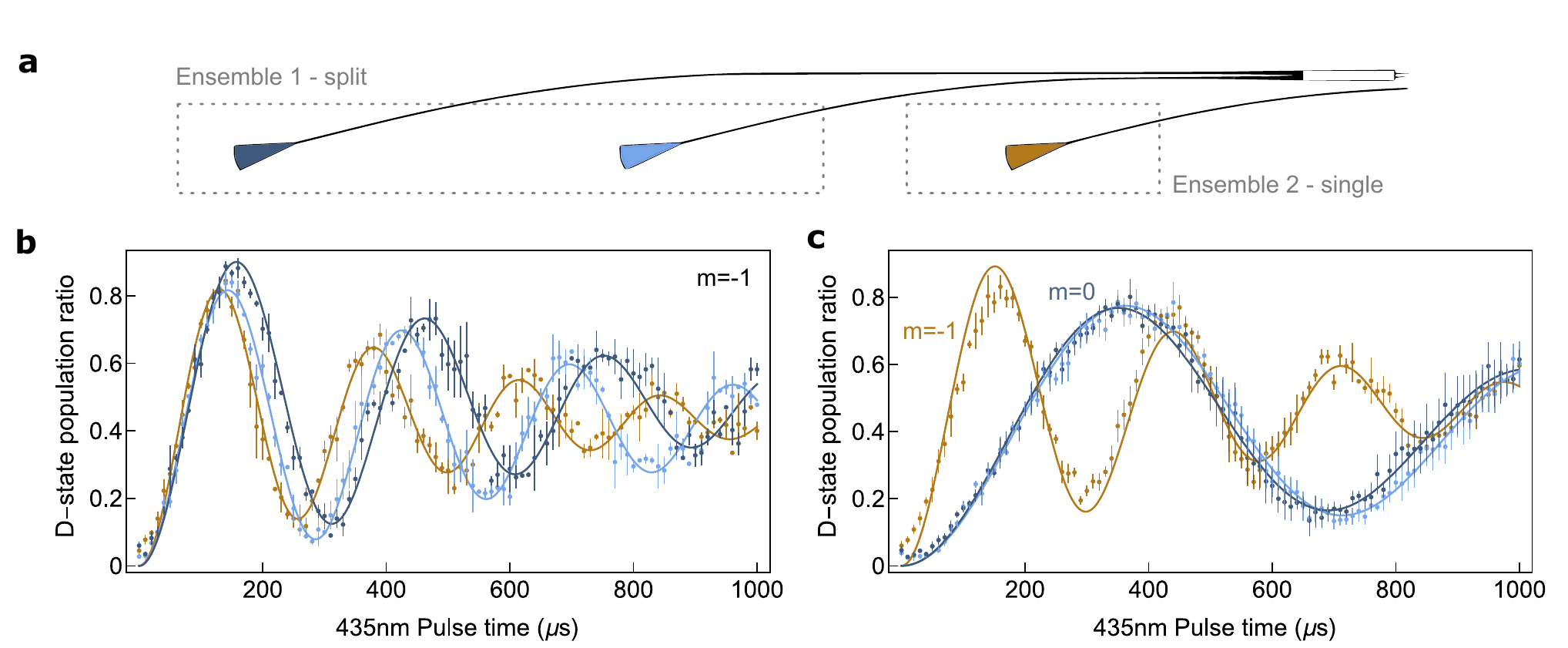}
    \caption{\textbf{Coherent qubit operations}.
    \textbf{a,} Schematic of split and single waveguide output grating couplers. Blue output grating couplers are routed from the splitter and share the same input, whereas orange is independent.
    \textbf{b,} Simultaneous Rabi flopping measurements of the three waveguide-illuminated ions. The trace colors correspond to the similarly colored gratings in \textbf{a.}
    The average ion population of the D-state ($\mF=-1$) was measured while scanning the 435 nm pulse time. The Rabi flops were fit with a single-ion decoherence Rabi oscillation formula in the Lamb-Dicke regime \cite{Semenin:2022OLP} (see Methods). \textbf{c,} Same as \textbf{b} but with the 435 nm light tuned to the ($\mF=0$) state for the two ion ensemble (light and dark blue), while the 435 nm light for the single ion (orange) remained tuned to the $\mF=-1$ state. Each data point was the average of three trials of 200 measurements of state-detection for each trace. Error bars show the standard error of the mean.}
\label{FIG:3rabi}
\end{figure*}

With the side-imaging system, the objective lens collected and focused light onto either a CCD camera (Andor Luca) or a 32 channel, linear PMT array (Hamamatsu H11659-200) with a multi-pinhole filter for the three waveguide sites. 
The 100\mic\! pinholes spatially filtered the ion fluorescence from background scattering to improve the signal-to-noise ratio (SNR) on the PMTs. Ultimately this arrangement achieved SNRs between 5 and 10 while supporting detection times on the order of milliseconds (4 ms for the measurement in Fig.\ \ref{FIG:3rabi}).
Fig.\ \ref{FIG:waveguidedetail}\textbf{a} shows fluorescence images of the ions illuminated by both free-space and waveguide beams. Each waveguide-illuminated ion, labeled (i) to (iii), was separated by a distance of 200\mic and placed in a local potential minimum created by the surrounding electrodes. To trap multiple ions simultaneously, we loaded each ion sequentially. The first ion was loaded at the loading hole position (marked with a red circle in Fig.\ \ref{FIG:waveguidedetail}\textbf{a}) using free-space beams (Doppler and repump) with a single-site potential solution (see Methods). This loaded ion was then shuttled towards the far-right position (iii) by applying the appropriate voltage waveforms to the interior control electrodes. This shuttling process was carried out without the use of cooling beams between sites, but once the ion reached the target position it was cooled by light from the output gratings. We subsequently loaded new ions by applying an additional potential minimum and repeating this process until three ions were stored at locations (i) to (iii) (see Supplementary Fig. 2).

While the Rabi flopping measurements in this paper used PMTs, we also fabricated SPADs for each ion-trapping site for future integrated detection. We shuttled and then detected ions with the SPADs using free space lasers, but ultimately could not operate the waveguides and SPADs simultaneously due to scattering from the waveguides that dominated the count rate.

The applied control and RF voltages produced an axial frequency of $2\pi \times 1.02$ MHz and radial frequency of $2\pi \times 3.52$ MHz, corresponding to a radial trap depth of 71 meV. 
Positioning the DC electrodes inside the RF rails led to a significant increase in DC electrode efficacy, such that static voltages within $\pm1$V were sufficient for achieving a 1 MHz axial frequency.

The 369 nm and 935 nm beams were used for cooling, state preparation, and detection of the ions, while the 435 nm beam was used for performing coherent operations on the E2 transition. Additional beams required for ion loading (399 nm and 393 nm) were delivered via free-space optics. This was done to reduce trap complexity, but future devices could easily incorporate waveguides that support photoionization.
Waveguides for delivering 760 nm repump light from the fourth output coupler were fabricated but never used. This wavelength can be used for depopulating the F state when a background gas collision transitions the ion to this state, but it was not necessary for these experiments given the several-hour lifetime achieved at the ultra-high vacuum pressures in the chamber.

All waveguide output gratings were designed to emit beams that overlapped at the trapping site 50\mic above the surface. While the 369 nm and 435 nm gratings were designed to focus at the ion's location, the repump beams were intentionally unfocused in order to ensure they overlapped with the ion position. Fig.\ \ref{FIG:waveguidedetail}\textbf{b} shows the measured waveguide profiles for 369 nm and 435 nm in $\hat{\textrm{\textbf{x}}}$ and $\hat{\textrm{\textbf{z}}}$.
The height of the ion (z) is controlled by adding an electric shim field in $\hat{\textrm{\textbf{z}}}$, with the resulting distances measured using the side-imaging camera.

As shown in Fig.\ \ref{FIG:waveguidedetail}\textbf{c}, the two beam profiles have closely overlapping foci near the 50\mic ion height, and near the center of the integrated SPADs.
The 369 nm integrated beam profile in Fig.\ \ref{FIG:waveguidedetail}\textbf{c} is measured using ion fluorescence well below saturation intensity, while ensuring that the signal is not cut off by the side-imaging pinholes. For the 435 nm quadrupole beam, the beam profile was characterized by measuring the Rabi frequency as a function of ion position. 
In both cases the profiles were fit to a Gaussian, showing a full-width-half-maximum beam width of 5.26\mic and 5.25\mic respectively. 

The in-situ measurements of the integrated beams agreed well with independent microscope measurements. Fig.\ \ref{FIG:waveguidedetail}\textbf{d} shows the theoretically simulated waveguide output coupler profile for 369 nm and 435 nm beams, where the beams are overlapping around 50\mic as designed. The $x^{\prime}$-axis in this figure is not the original $x$-axis but instead corresponds to the projection of the $x$-axis on a plane that is perpendicular to the trap surface and runs through the centers of the gratings.

\subsection{Simultaneous multi-ion/multi-state addressing}

In these experiments, ensembles consisting of 2$+$1 trapped ions (2 ions with shared waveguide inputs and a single ion with independent control) were probed simultaneously. Each ion was first optically pumped to $\sstate$ and then a 435 nm laser resonant with the quadrupole transition was used to demonstrate Rabi flopping with $\dstate$ (Fig.\ \ref{FIG:3rabi}).
The ion positions for the split beams (blue traces) were set to maximize the interaction strength for each ion within each 435 nm beam. Ideally, this would correspond to the maximum intensity, but other factors like micromotion could slightly offset these optima. 
The two ions exhibited similar Rabi rates ($\delta \Omega = 0.064(4) \Omega$), indicating a near 50/50 split for the MMI. Though not necessary in this case, the interaction strengths could have been fine-tuned by adjusting the position of each ion to equalize their Rabi rates. The orange data indicates the independently addressed ion. Each data set was fit with a single-ion, two-state transition model that accounted for the average motional quanta $\bar{n}$ of the ion in the Lamb-Dicke regime.

In practical realizations of quantum computers, quantum sensors, or atomic clocks, it may be beneficial to probe multiple transitions with different sensitivities in order to calibrate magnetic fields. To demonstrate this capability, we simultaneously probed two different transitions (one with each ensemble) as shown in Fig.\ \ref{FIG:3rabi}\textbf{c}. This measurement was similar to Fig.\ \ref{FIG:3rabi}\textbf{b}, but with a different optical qubit frequency used for the ions addressed by the split waveguides. These ions (dark and light blue) transitioned from $\sz$ to $\dz$ (the \negmedspace\ybion\medspace clock transition), while the single-waveguide ion transitioned from $\sz$ to $\dm$. Due to their different coupling coefficients, they exhibited different Rabi rates. The frequency required for the $\delta m=0$ transition state is first-order insensitive to magnetic fields and is naturally well-overlapped (the difference in Rabi rates is $\delta \Omega = 0.039(4) \Omega$). This further confirmed the near 50/50 power split performance of the MMI splitter, as the Rabi rates became closer when the transition was not susceptible to spatial magnetic field gradients.

The ions were not cooled to the motional ground state in this experiment ($\bar{n} \approx 30$, based on previous Rabi fits), which leads to decay in the Rabi contrast as pulse time increases \cite{wineland:1998}. This $\bar{n}$ is higher than the Doppler limit ($\approx$10 quanta), primarily due to insufficient cooling times (discovered later in the experiment) and less optimal Doppler cooling caused by the fixed polarization and $k$-vector of the light emitted from the output coupler relative to the applied magnetic field, based on comparisons to the lower $\bar{n}$ measured with free space beams. Heating rates were measured at site (ii) to be 1-3 q/ms (see Methods).
The detection fidelity achieved in this experiment is lower than typical, primarily due to collection efficiency limitations of the side-imaging system and the higher $\bar{n}$. Our chamber setup unavoidably increased the distance between the side-imaging objective lens and the trapped ion, making it challenging to use high numerical aperture objective lenses. Additionally, while greatly improved by using side-imaging, there remained some scattering from the waveguide output couplers that contributed to background noise.

\begin{figure}[t!]
\centering
\includegraphics[width=0.98\columnwidth]{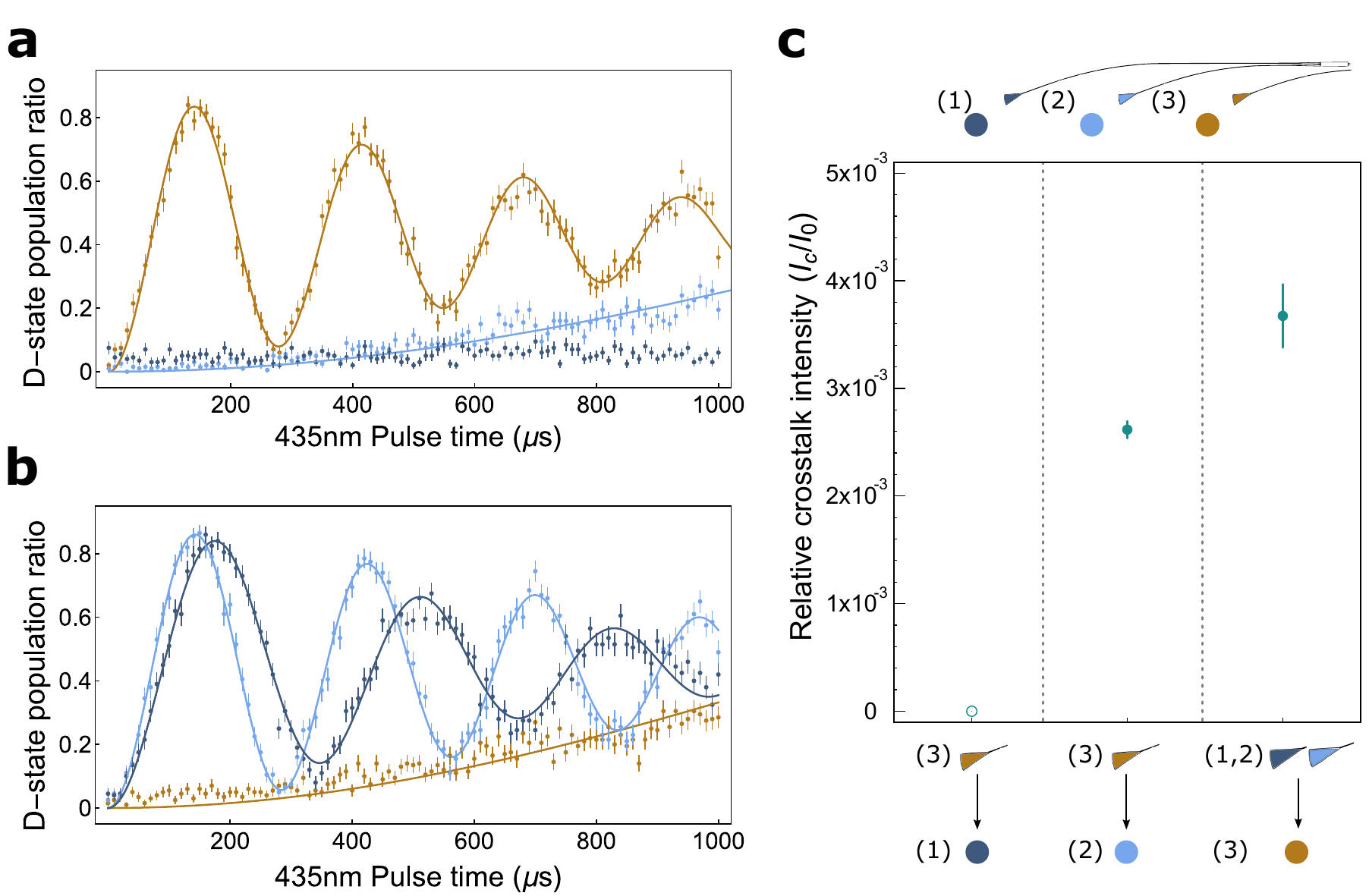}
    \caption{\textbf{Optical cross-talk}
    \textbf{a,} Measured cross-talk between the single ion ensemble (orange) and two ion ensemble (light and dark blue). Rabi oscillations with the D-state ($m_F=-1$) were measured while addressing the single ion ensemble (orange). Cross-talk was evident in the slow transition rate of the nearer ion (light blue) in the two ion ensemble (light and dark blue), and was more attenuated for the more distant ion.
   \textbf{b,} A similar measurement as \textbf{a}, but instead addressing the two ion ensemble (light blue and dark blue). Cross-talk was evident in the slow transition rate of the single ion ensemble (orange). Error bars show the standard error.
   \textbf{c,} Relative cross-talk intensities of transition light. Ratios were calculated from fits of Rabi frequencies and plotted for each trap site.}
\label{FIG:cross-talk}
\end{figure}

\subsection{Cross-talk characterization}

An important consideration in any multi-ion system is optical cross-talk, as the unintended light on neighboring ions introduces errors in the clock or qubit. Cross-talk errors for integrated addressing of a single ion in a chain have been measured as a function of displacement from the center of the beam profile \cite{mehta:2016}, as well as in a multi-zone trap \cite{mehta:2020}.

To measure cross-talk, light was applied at one site while Rabi transitions were measured at all sites.
The ion(s) of a single ensemble were probed using the $\sz$ to $\dm$ transition, in the same way as Fig.\ \ref{FIG:3rabi}\textbf{b}, while the corresponding optical cross-talk on the ion(s) of the other ensemble was observed. Fig.\ \ref{FIG:cross-talk}\textbf{a} shows that addressing the single ion ensemble resulted in a Rabi frequency $\Omega_0$ for the single ion and a cross-talk-induced Rabi frequency $\Omega_{c}\sim0.05 \Omega_0$ on the closest ion of the other ensemble. This corresponds to a relative cross-talk intensity of $I_{c}/I_{0}=0.0026(1)$, where $I_{0} \propto \Omega_0^2$ is the driven intensity and $I_{c}$ is the cross-talk induced intensity. 

It is notable that only ion (ii) of the ensemble showed measurable optical cross-talk. If scattering due to waveguide edge roughness caused significant waveguide-to-waveguide coupling between single and ensemble waveguides, cross-talk would be similar for ion (i) and (ii). As shown in Fig.\ \ref{FIG:cross-talk}\textbf{a} (dark blue), cross-talk was proximity dependent and therefore not significantly caused by the input couplers or waveguide-to-waveguide coupling on chip, but rather was dominated by scattered light from the output coupler. Another experiment \cite{mehta:2020} measured lower cross-talk (-60 dB) with gratings that emitted 729 nm light in the -1 order mode, which suppresses higher order modes. Given our lithography tolerances and a shorter 435 nm wavelength, we were limited to designing gratings that emit in the +1 order mode, and subsequently generated higher order modes that resulted in greater optical crosstalk. In addition, shorter wavelengths are more sensitive to fabrication imperfections that can lead to scattering.

Fig.\ \ref{FIG:cross-talk}\textbf{b} shows the data for the opposite case, where the two ion ensemble was addressed. The single ion measured a relative cross-talk intensity of $I_{c}/I_{0}=0.0036(2)$. The two measurements of cross-talk have similar intensities which suggests that the source of the cross-talk has little directional dependence.

\subsection{Toward a fully-integrated system}

Our trap was equipped with SPADs at each trap site for integrated single-site detection. The SPADs integrated in this trap were the same as previously designed \cite{setzer:2021}, however an aluminum mesh was added to shield the ion from the electric field of the SPAD (see Methods). This change allowed for trapping \ybion\ directly above the SPAD while it was active. 
We operated one SPAD independently and measured a count rate of over $7$ kcps for one quadrant of the SPAD, with dark counts below $1$ kcps. These results are better than the previous SPAD performance reported for $^{174}\textrm{Yb}^+$ \cite{setzer:2021}.
Unfortunately, scattered light originating from the waveguides and output couplers propagated through the various layers of the device, saturating the SPADs and prohibiting integrated state-readout. In the future, features such as light baffles could mitigate this scatter and enable simultaneous fully integrated addressing and state-readout.

\section{Discussion} 
In this work, we demonstrated the individual addressing of trapped-ions at multiple sites on a surface trap using MMI waveguide splitters and fully integrated multi-color waveguides for cooling, state preparation, coherent operations and detection. This combination of integrated waveguide delivery and multi-ion control with single channel optical inputs shows a path towards scaling trapped-ion systems like quantum computers, where potentially millions of qubits \cite{reiher:2017} will be needed for a truly useful general-purpose quantum computer. 

A more near-term application of this work is to create a robust and portable optical ion clock by replacing the optomechanics in lab-scale systems with integrated photonics like lasers-on-chip \cite{Elshaari:2020Npho}, modulators \cite{hogle:2023}, detectors \cite{setzer:2021, Reens:2022PRL}, and the waveguides and splitters described here. 
They can also be used to improve the clock accuracy by supporting multiple ensembles of ions that can be concurrently probed with different interrogation times, and by eliminating deadtime \cite{Schioppo:2017NPho} with ensembles that have non-overlapping detection windows. Clock performance can be further enhanced by probing systematics \cite{Roseband:2013arx,Kim:2022NP} while simultaneously operating the clock.

\section{\bf {\centerline{METHODS}}}

\paragraph{\bf {\centerline{Trap design and fabrication}}}

\subparagraph{\it }

This trap was designed and fabricated at Sandia National Laboratories and shares features common to previous traps with integrated photonics \cite{ivory:2021,setzer:2021}, notably waveguides and MMI splitters for optical addressing as well as SPADs for state readout. 
The SPADs could not be operated when light was delivered via integrated waveguides, as the scatter from those components overwhelmed the ion signal on the SPADs. 
We implemented a grounded metal mesh above the SPAD to reduce unpredictable charging effects \cite{ivory:2021} and consequent shuttling problems. It serves a similar function as an optically transparent and electrically conductive coating (e.g. indium tin oxide).
The mesh allowed the ion to be transported over the entire aperture of the SPAD while the SPAD was active, which was not possible in previous Sandia-fabricated SPAD traps \cite{setzer:2021}. The mesh was designed to be  1\mic thick with many square 3\mic $\times$ 3\mic openings to allow ion fluorescence to pass through. (see Supplementary Fig.\ 1). The mesh transmits roughly 55\% of the ion fluorescence to the SPAD, and based on electrostatic simulations reduces the electric field at the ion by 16 dB compared to not having a screen.

\subparagraph{\bf {\centerline{Experimental system}}}

\subparagraph{\it Ion trapping.}

To avoid coating the electrodes with ytterbium, the trap has several loading holes that penetrate the entire thickness of the trap and its substrate.
Ytterbium metal in a stainless steel tube is heated with current to vaporize atoms in the direction of the loading holes. The ytterbium atoms are ionized using a two-photon process with laser beams at 393 nm and 399 nm, at which point ytterbium ions are electrically confined by the RF and DC fields created by the trap electrodes.

The trap potential well is positioned 50\mic above the chip's surface. In the loading zone ions are illuminated and cooled using free-space Doppler (369 nm) and repump (935 nm) beams. 
Additional sidebands of 2.1 GHz and 14.7 GHz are applied to the Doppler beam using electro-optic modulators (EOMs) for optical pumping and Doppler cooling.
During the optical pumping sequence, the ion is prepared in the $\sz$ state by applying only the extra 2.1 GHz frequency to the standard Doppler frequency. This preparation can be observed through a modified detection step that only detects atoms in the $F=1$ state. 

The motional heating rate $\dot{\bar{n}}$ of the axial trap mode was measured for an ion at site (ii) illuminated with both free-space and waveguide beams.
For the free-space case, we used sideband thermometry \cite{roos:2000} on the $\sz$ to $\dm$ transition, and measured a heating rate of $\dot{\bar{n}}$ = $2.71\pm1.63$ q/ms for an ion that was initially cooled to $\bar{n} \sim 2$.
Using waveguides, we measured a heating rate using the Rabi flopping method described below.
In this case a free-space beam was used only for detection to allow for top-imaging. This enabled more accurate measurements due to the higher fidelity detection and shorter detection times, while still using the integrated waveguides for all other light delivery. As presented in Supplementary Fig.\ 3, we measured $\dot{\bar{n}}=1.25\pm0.14$ q/ms, comparable to the case without any integrated light delivery.

\subparagraph{\it Optical losses in waveguides.}
Each waveguide on the device has both an edge coupler and grating coupler input that are combined on chip, though only the grating couplers were used for this work. The 369 nm and 435 nm waveguides were separately tested and a total power reduction of $\approx$20 dB for each was measured. This includes the power reduction from the input combiner (3 dB) and splitter (3 dB) for separate sites, as well as a total loss of $\approx$14 dB. This total loss is the sum of the estimated insertion loss at the input grating (6-7 dB), waveguide propagation loss (1.35 dB/cm at 369 nm and 0.9 dB/cm at 435 nm over 5 mm), and the simulated output grating efficiency (6-7 dB). While higher power reductions were measured on some devices, the optically good ones were consistently in the 20 dB range.

In experiments with ions, a total loss of 49 dB for the 435 nm light was calculated based on the input power and Rabi rate of the ion, significantly higher than the $\approx$14 dB ideal total loss measured outside the vacuum chamber. A 369 nm waveguide on the actual device used in the experiments measured $\approx$14 dB total loss prior to experiments, but the 435 nm waveguide was not measured, and may have had significantly higher loss due to fabrication imperfections.
In the experiment each laser beam is launched 75 mm from the grating outside the vacuum chamber and passes through a viewport at a $45^{\circ}$ angle, causing significant beam distortion that may contribute to the different total loss. Future systems will employ edge coupling to improve coupling efficiency, as demonstrated in other experiments \cite{mehta:2020} and described below.

\subparagraph{\it Waveguide scalability.}
In addition to the $1 \times 2$ splitter used in this paper, other splitting ratios are possible by varying the interaction length in the MMI splitters, as are other fanout ratios ($1 \times$N). Larger values of N reduce the number of required optical launches to the chip but increase the amount of power in the input waveguide. This is generally not an issue for visible wavelengths; a demonstration of 300 mW of 729 nm light in a silicon nitride waveguide \cite{mehta:2020} shows that many sites could be supported from a single launch. However, UV wavelengths tend to cause greater damage, so the development of low-loss waveguides that do not degrade from photodarkening with UV powers up to $\sim$100 mW is critical to realizing this approach to scaling.

To show that much better coupling is possible for UV wavelengths, we edge-coupled 369 nm light from a lensed fiber to a test chip and back into a separate lensed fiber.  The test chip used the same waveguide material and structure as the trap. Assuming equal coupling losses on and off chip and subtracting for the measured waveguide absorption loss (0.6 dB), we measured a 4.1 dB coupling loss from the fiber to the chip.
Over short time periods the total power loss remained stable for in-waveguide powers ranging from 7 mW to 24 mW, but the power degraded over longer (hour) time periods.  Multiple causes for this degradation are possible, including surface contamination at the fiber or edge faces, absorption in the inverse taper used for coupling, or photodarkening in the waveguide itself. Future research will be conducted to understand this behavior.

\subparagraph{\it Side-imaging system.}
One of the main difficulties in integrated waveguide light delivery schemes is detection, since scattered photons from the grating output couplers are at the same wavelength as the ion fluorescence and cannot be filtered.
To address this a side-imaging technique was developed. The fluorescence from an ion was collected with a 2-inch diameter objective lens from the side of the chamber, and guided to either a CCD camera or linear PMT array. 
The bottom half of photons emitted by the ion are obscured by the chip, which gives an effective NA=0.12, compared to the conventional top imaging system with NA=0.28.

A 1D PMT was used for simultaneous independent detection of multiple ions. The ion separation in the image plane was comparable to the cell spacing of the PMT array so that each ion was detected by separate consecutive PMT pixels.
To exclude scattered light from the trap, a pinhole array with 100\mic diameter holes was placed in front of the PMT array.

\subparagraph{\it Generating multi-ion trapping solutions for multiple locations.} 

The DC voltage solutions for most ion trap experiments are created by constraining the electric field to be zero along three axes and the derivative of the electric field to have a specific positive value along one axis, all at the single point where we wish to trap the ion. This defines a potential well with a specific motional frequency. Since this problem is underconstrained and there are many possible solutions, other requirements (e.g. voltage limits on the electrodes) and desirements (e.g. applying voltages to fewer electrodes) are applied.
Small changes to compensate the ion for deviations in the local electric field were computed separately and applied additively. 

While this system served well for many uses, the need to precisely position multiple ions with respect to fixed laser beams, as well as the need to shuttle newly loaded ions into position, required a more flexible solution. Initially a strategy of pre-computing all combinations of positions for all wells was attempted.
Since the number of possible combinations of well positions grows exponentially with the number of ions, so did our voltage solutions. This quickly became infeasible. The resolution to this problem was to create these solutions in real-time. For the real-time multi-ion solution the constraints were specified then combined to form a linear system that could be solved.
The desired positions of all ions are provided as an input, along with the compensation fields at those positions. The system looks up the appropriate constraint matrices, concatenates them, and solves the resulting system. 

\subparagraph{\bf {\centerline{Theoretical approaches}}}

\subparagraph{\it Rabi flopping in the Lamb-Dicke regime}

To quantify the average number of motional quanta in the system, we fit the experimentally obtained Rabi flop data to a dephasing model.

First, we check if our system is in Lamb-Dicke regime ($\sqrt{\bar{n}_{k}}\eta_{k} \ll 1$).
Here the Lamb-Dicke parameter $\eta_{k}$ is defined as $\eta_{k} = \frac{2\pi}{\lambda} \cos{\theta} \sqrt{\hbar/(2 m \omega_{k})}$
where $\lambda=435$ nm is the wavelength, $m$ is the mass of \ybion, $\hbar$ is plank's constant, $\theta$ is the beam angle, and $\omega_{k}$ is the secular motional frequency. With the given parameters, we confirmed that $\sqrt{\bar{n}_{k}}\eta_{k} \ll 1$ if $\bar{n}<50$, which is valid in our experiment.

Because the depths of ion traps are much larger than the thermal energy of the ion, we fit the Rabi carrier function in a Lamb-Dicke regime as $P(t)$, the total population of excited state, as follows \cite{Semenin:2022OLP}:

\begin{equation}
\begin{aligned}
P(t)=
\frac{a}{2} \left( 1-\frac{ f_{1}(t)  }{ f_{2}(t)} \right)
\end{aligned}
\end{equation}
where
\begin{equation}
\begin{aligned}
f_{1}(t)=
Re[e^{i\Omega t}] \prod_{k=1}^{N}  e^{-i \Omega {\eta^2_{k}} t/2 } \left( 1- \frac{\bar{n} e^{i \Omega {\eta^2_{k}} t}}{\bar{n}+1} \right)
\end{aligned}
\end{equation}
and
\begin{equation}
\begin{aligned}
f_{2}(t)=
\prod_{k=1}^{N} \left( (\bar{n}+1)-2\bar{n} \cos{(\Omega_0 \eta^2_{k} t)} + \frac{\bar{n}^2}{\bar{n}+1} \right),
\end{aligned}
\end{equation}
for generalized $N$ normal vibrational modes, characterized by $N$ secular frequencies $\omega_k$. Here $a$ is the initial population of the ground state, $\Omega_0$ is the Rabi frequency of the ion at rest, and $\eta_k$ is the Lamb-Dicke parameter for the given ion in the $k$th mode.

\subparagraph{\it Cross-talk - evanescent coupling calculation}
We estimated an evanescent coupling between waveguides, assuming two parallel waveguides of the symmetric and anti-symmetric mode. The coupling length was calculated by considering the difference in refractive index between two coupled modes whose refractive index is around $n \sim 1.51$. For the parameters in our device, where the waveguides are separated by $6.26$ \mic over a distance of $1.7$ mm, the evanescent coupling is on the order of $10^{-25}$, which is negligible. This analysis does not include fabrication imperfections like waveguide edge roughness that can increase the waveguide-to-waveguide cross-talk.

\paragraph{\bf {Data availability}}
All relevant data are available from the corresponding authors upon reasonable request.

\bibliographystyle{naturemag}

\newpage

\paragraph{\bf {Acknowledgements.}}
We thank Justin T. Schultz for his contribution in the early stages of side-imaging system setup and Tharon D. Morrison for the waveguide UV tolerance test.
We thank the members of Sandia's Microsystems and Engineering Sciences Application (MESA) facility for their fabrication expertise and for helpful comments on the manuscript. This work was supported by the Defense Advanced Research Projects Activity (DARPA). Sandia National Laboratories is a multi-mission laboratory managed and operated by National Technology and Engineering Solutions of Sandia, LLC, a wholly owned subsidiary of Honeywell International Inc., for the U.S. Department of Energy's National Nuclear Security Administration under Contract No. DE-NA0003525. This paper describes objective technical results and analysis. Any subjective views or opinions that might be expressed in the paper do not necessarily represent the views of the U.S. Department of Energy or the United States Government.

\paragraph{\bf {Author Contributions.}}
J.K., W.J.S., D.S. and H.J.M. conceived the experiment. J.K. and W.J.S. took the measurement and analyzed data. M.G. designed the integrated optics while also performing waveguide tests and simulations with N.K. J.V.D.W. developed multi-ion solution. R.L. setup the Doppler laser system. M.B. oversaw the trap fabrication. 
J.K., W.J.S., D.S. and H.J.M. wrote the original manuscript. 
All the authors reviewed and edited the manuscript.
H.J.M. supervised the project. 

\paragraph{\bf {Competing Interests.}}
J.K., W.J.S., and H.J.M. are inventors in patent applications related to the "side-imaging" technique used to collect ion fluorescence. The remaining authors declare no competing interests.

\newpage

\section{Supplementary figures}

\begin{figure*}[t!]
\centering
    \includegraphics[width=1.0\columnwidth]{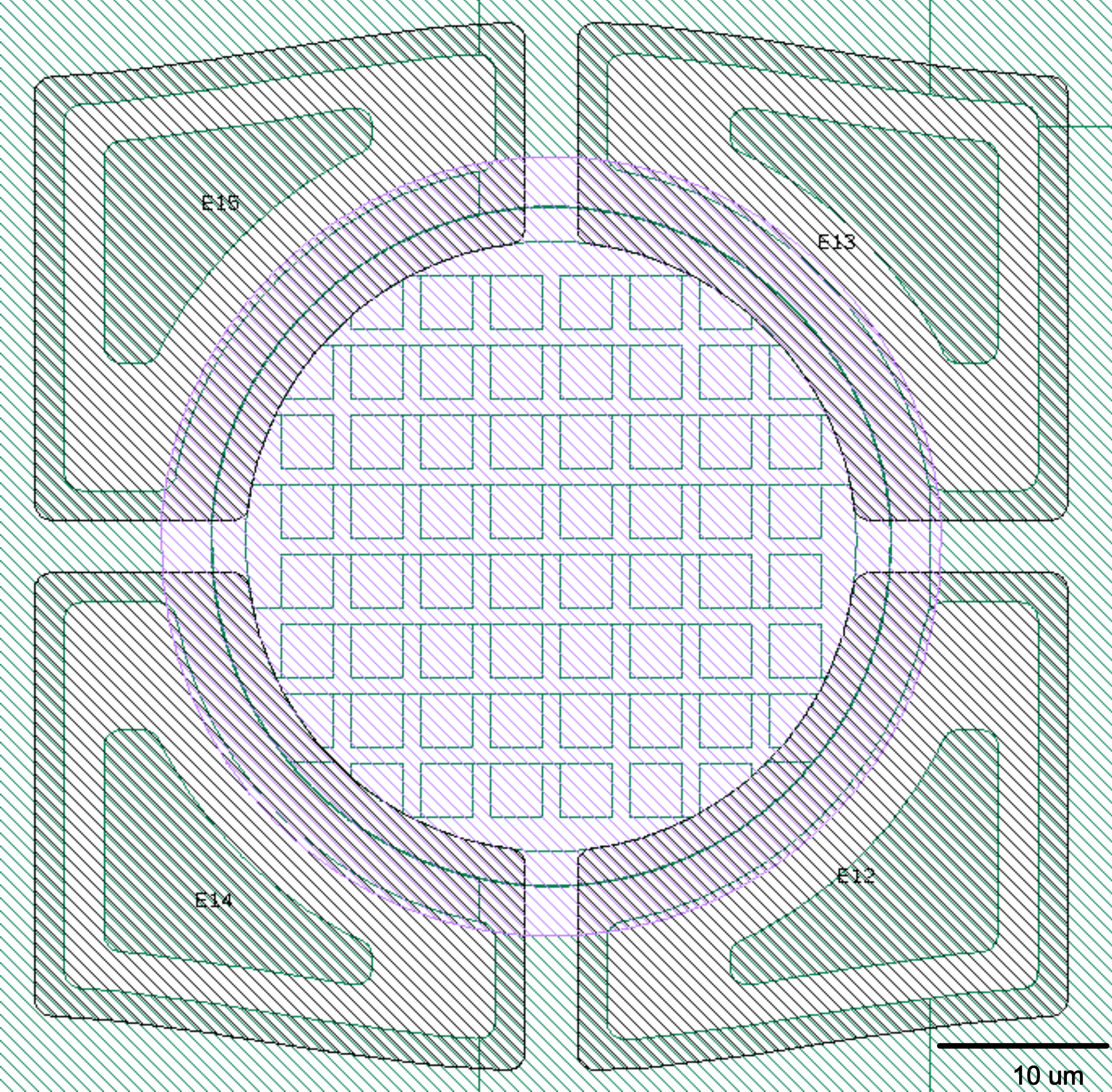}
    \caption{\textbf{Lithography mask in the SPAD region}
    A metal mesh is located on top of the SPAD (purple circle), consisting of 3\mic $\times$ 3\mic holes with 1\mic wide metal traces. The mesh and SPAD are surrounded by nearby electrodes, also shown in Fig.\ \ref{FIG:SystemIntro}\textbf{a}.
  }
\label{FIG:sup_spad}
\end{figure*}

\begin{figure*}[t!]
\centering
    \includegraphics[width=1.9\columnwidth]{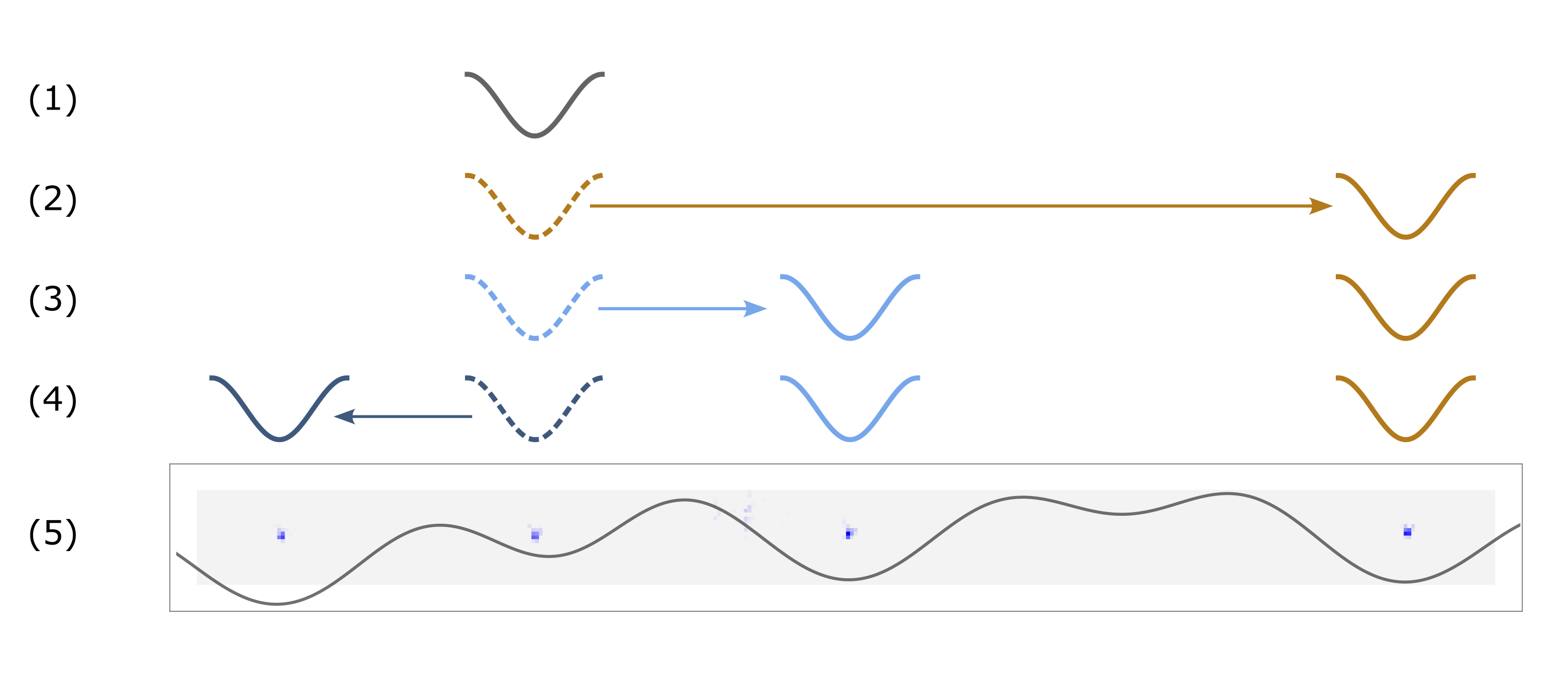}
    \caption{\textbf{Multi-ion loading sequence} 
    Three waveguide-illuminated independent ions are loading sequentially, starting from one loading location that used free-space beams. 3+1 ions are trapped simultaneously by generating potential wells and loading/shuttling ions via steps (1) to (5). The potential plot at (5) represents a real four-ion solution that is used in this work, providing secular frequencies of $2\pi \times 1.02$MHz.
  }
\label{FIG:sup_loading}
\end{figure*}

\begin{figure*}[t!]
\centering
    \includegraphics[width=1.5\columnwidth]{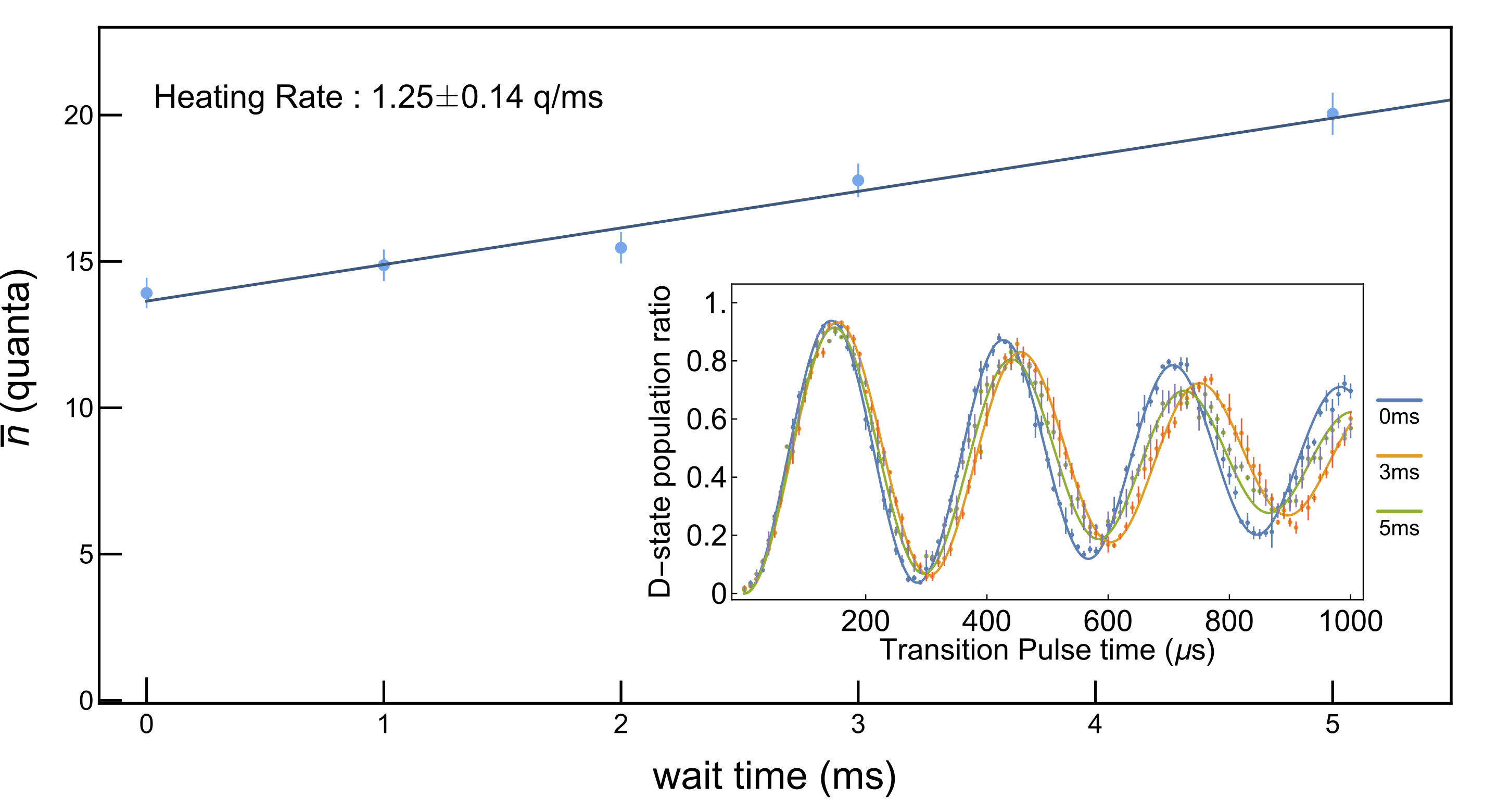}
    \caption{\textbf{Heating rate measurement} 
    A heating rate of $1.25 \pm 0.14$ q/ms is measured using fully integrated beams for state preparation, cooling, and Rabi flopping, while a free-space beam is used for detection. \textbf{inset:} Measured Rabi flopping for different wait times.
  }
\label{FIG:sup_HR}
\end{figure*}

\end{document}